\pdfoutput=1

\documentclass[11pt]{article}

\usepackage[final]{acl}

\usepackage{times}
\usepackage{latexsym}

\usepackage[T1]{fontenc}

\usepackage[utf8]{inputenc}

\usepackage{microtype}

\usepackage{inconsolata}

\usepackage{graphicx}

 \usepackage{booktabs}
 \usepackage{multirow}
 \usepackage{multicol}
\usepackage{hyperref}
\usepackage{tabularx}

\usepackage{listings}
\usepackage{xcolor}

\lstdefinestyle{prompt}{
  basicstyle=\ttfamily\small,
  backgroundcolor=\color{gray!10},
  frame=single,
  breaklines=true,
  showstringspaces=false
}
\title{TDFlow: Agentic Workflows for Test Driven Development}

\author{
    \textbf{Kevin Han\textsuperscript{1}},
    \textbf{Siddharth Maddikayala\textsuperscript{2}},
    \textbf{Tim Knappe\textsuperscript{1}},
    \textbf{Om Patel\textsuperscript{1}},\\
    \textbf{Austen Liao\textsuperscript{3}},
    \textbf{Amir Barati Farimani\textsuperscript{1}},
    \\
    \\
    \textsuperscript{1}Carnegie Mellon University,
    \textsuperscript{2}UC San Diego,
    \textsuperscript{3}Johns Hopkins University, \\
    \texttt{\{kevinhan, barati\}@cmu.edu}
}

\begin{document}
\maketitle
\begin{abstract}
We introduce TDFlow, a novel test-driven agentic workflow that frames repository-scale software engineering as a test-resolution task, specifically designed to solve human-written tests. Given a set of tests, TDFlow repeatedly proposes, revises, and debugs repository-scale patches using precisely engineered sub-agents and tightly constrained tools. The workflow decomposes software engineering program repair into four components governed by respective sub-agents. This simple, forced decoupling of patch proposing, debugging, patch revision, and optional test generation (1) reduces long-context burden on any individual sub-agent, (2) focuses each sub-agent on specific, pre-defined sub-tasks, and (3) allows for specialized performance improvement on specific sub-tasks. When provided human-written tests, TDFlow attains 88.8\% pass rate on SWE-Bench Lite (an absolute improvement of 27.8\% over the next best baseline) and 94.3\% on SWE-Bench Verified. Manual inspection of the 800 TDFlow runs within SWE-Bench Lite and Verified uncover only 7 instances of test hacking, which were subsequently counted as failures. Furthermore, we show that the primary obstacle to human-level software engineering performance lies within writing successful reproduction tests. We envision a human-LLM interactive system powered by TDFlow where human developers write tests solved by LLM systems. Together, these results indicate that modern LLMs, when embedded in a narrowly engineered, test-driven workflow, already achieve human-level test resolution -- with the final frontier for fully autonomous repository repair being the accurate generation of valid reproduction tests.
\end{abstract}

\section{Introduction}
Advancements in the code reasoning, long context, and agentic capabilities of large language models (LLMs) have led to the rise in LLM use cases as autonomous coders capable of fixing bugs at the repository scale \citep{liu2024large}. Models such as Claude 4 Sonnet, Kimi K2, Qwen3-Coder, and Gemini 2.5 Pro are provided with a set of file-interfacing tools and trained specifically on long-horizon reinforcement learning objectives \citep{anthropic2025claude4, team2025kimi, team2023gemini, hui2024qwen2}. During inference time, these models are equipped with the same set of tools, provided an issue description and access to a repository, and are tasked with writing test cases, thinking of edge cases, and proposing a patch that addresses the issue. Afterwards, the patch is run against a series of hidden tests to check if the issue has been addressed. Current repository-scale software engineering benchmarks are focused on solving all hidden tests when provided with an issue description \citep{jimenez2024swebench, swebenchlive, swebenchmultimodal, chowdhury2024swebenchverified}. However, in many real-world settings, software engineering and bug resolution is guided by the ethos of test driven development (TDD), where unit and integration tests are written before the actual code itself is implemented. The benefits of TDD include higher code quality, reduced recurring bug rates, and facilitate clearer, modular code that is easier to maintain and extend \citep{TDDvsBDD1, AlSaqqa2020AgileSD}. On the other hand, the primary downside of TDD is the longer development time associated with writing tests for all features \citep{micrsoftTDD}. \textbf{We hypothesize that LLMs can streamline TDD by solving tests written by human developers}. We envision a human-LLM collaborative software engineering system in which human guidance is provided to an LLM system via supplying ground-truth tests while LLM systems are relied upon to narrowly solve the tests and thereby expand the functionality of the repository.

\begin{figure*}[!t]
  \centering
  \includegraphics[width=0.9\linewidth, trim=1.5cm 3cm 1.5cm 3cm, clip]{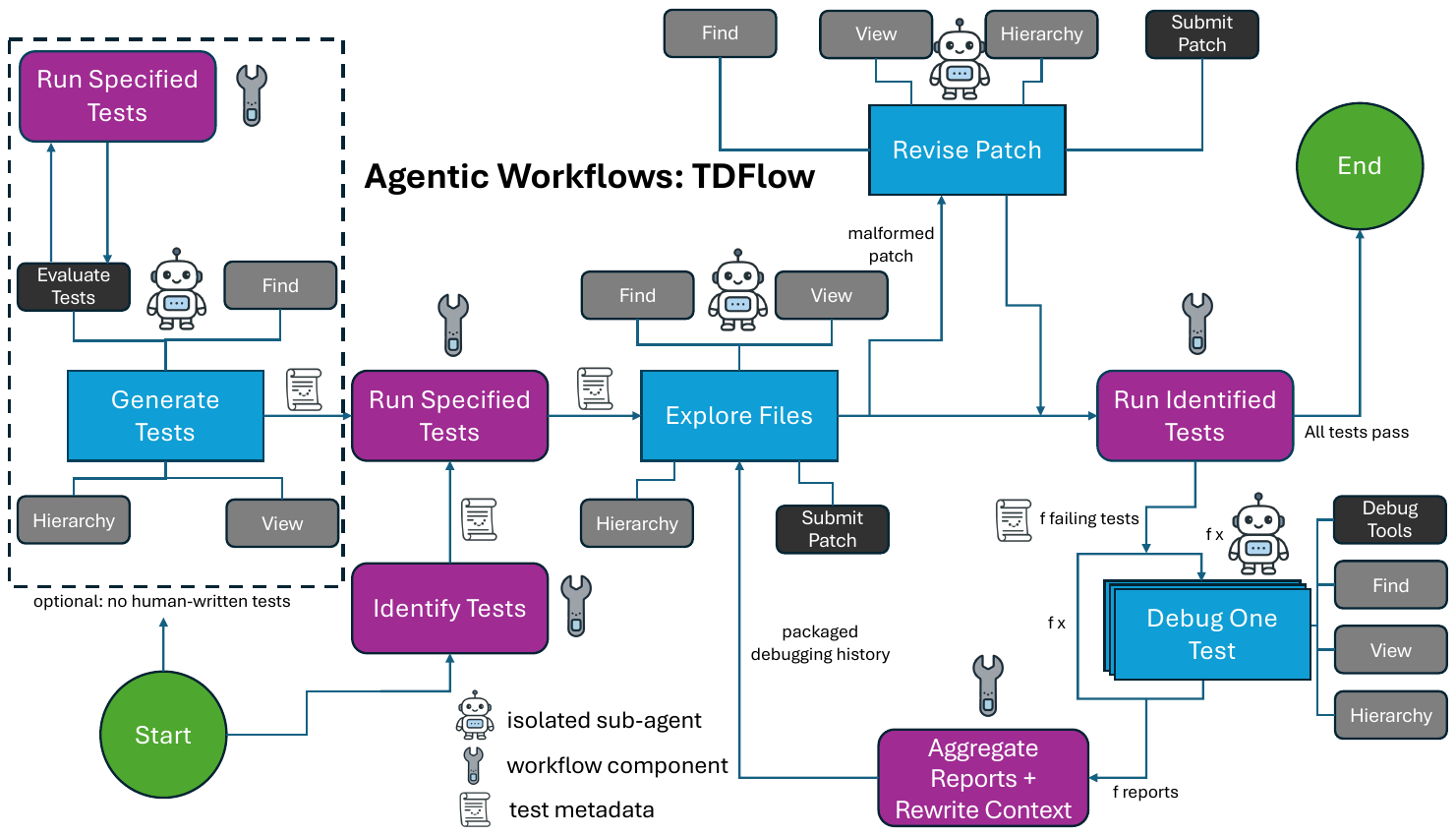}
  \caption {The workflow behind Test-Driven Flow (TDFlow). TDFlow is a purely test driven, agentic \textbf{workflow} for resolving repository scale issues. The entrypoint to TDFlow begins with either human-written reproduction tests or, optionally, to have TDFlow generate reproduction tests. Afterwards, the tests are run and provided to the \textit{Explore Files} LLM sub-agent with the sole task of exploring the repository in order to propose a patch. The tests are run on the proposed patch before the \textit{Debug One} sub-agent debugs each failing test individually with a dedicated debugger tool and generates reports. Those reports are used by the \textit{Explore Files} sub-agent to propose another patch.}
  \label{fig:overview}
\end{figure*}

In this work, we develop \textbf{T}est \textbf{D}riven Agentic Work\textbf{flow} (TDFlow), an agentic workflow specifically designed to solve human-written, ground-truth test cases at the repository scale. TDFlow consists of several LLM sub-agents operating within a precisely context-engineered environment. In the context of solving ground-truth test cases, workflows consisting of sub-agents pose several advantages over a single model, general-purpose, monolithic agent framework equipped with a broad set of tools. Primarily, monolithic LLM agents, or agents comprised of a single model and are indivisible in feature and capability with respect to each subtask, struggle with simultaneously decomposing and executing complex tasks into smaller subtasks \citep{chen2025evaluating}. In TDD, once the ground-truth tests are written, the development workflow becomes 1) navigating the codebase to develop a solution to pass tests, 2) debugging any failed tests, and 3) repeating the process until all tests are resolved \citep{pairTDD}. A strict workflow can explicitly model and enforce this decomposition of tasks, allowing each LLM sub-agent to focus on executing well-defined subtasks without the burden of additional planning.

TDFlow is specifically tailored for resolving human-written test cases at the repository scale. On SWE-Bench Lite, when human-written, ground-truth tests are provided, the TDFlow system performs 27.8\% better than the previous systems agentic software engineering systems. When generating its own tests, TDFlow achieves 69.8\% performance on SWE-Bench Verified. However, when provided with human-written tests, TDFlow's success rate reaches 94.3\% -- suggesting that TDFlow has achieved human-level performance in test resolution and can already be used in a human-LLM collaborative software engineering paradigm where humans write test cases which are solved by LLMs.

We show that the final hurdle in human-level, fully autonomous software engineering with TDFlow lies within test generation, rather than issue resolution. Finally, we find a total of only 7 examples of test hacking after having three trained software engineers perform a manual analysis into all 300 SWE-Bench Lite and 500 SWE-Bench Verified human-written test runs.

\section{Related Work}

\subsection{Test Driven Development}
Test driven development (TDD) has been repeatedly determined to significantly improve code quality at the expense of increased development time. An empirical study on four high impact teams at IBM and Microsoft found that TDD improved code quality, defined by defect density, by up to 90\% but also increased development time by up to 35\% \citep{micrsoftTDD}. Another work split 24 professional engineers into 2 groups, one group developed with a TDD approach while the other group developed without a TDD approach \citep{pairTDD}. Results showed that TDD led to 18\% higher quality code defined by pass rate on functional black box tests. Similar to the findings of \citet{micrsoftTDD}, the authors discover that TDD required 16\% more development time. Overall, a large number of in-the-wild studies have been conducted both enforcing the efficacy as well as the additional time burden of TDD \citep{IBMTDD2, TDDvsBDD1, mishra2022comparative,Bakhtiary2020TheEO}. The studies also mention the additional time burden being a prohibitive factor preventing more teams from adopting a test first approach to software engineering. With TDFlow, the increase in development time can be mitigated by using LLM systems to solve the tests that are written by human developers, with humans then verifying the resulting patch proposed by the LLM.

\subsection{Software Engineering Agents and Benchmarks}
\textbf{SWE-Bench} is a benchmark consisting of 2,294 real world softare engineering problems scraped from Github issues and corresponding pull requests. Provided an issue description and repository, agents are tasked with proposing a patch to resolve the issue description. The success of the patch is determined by whether or not the patch passes a series of hidden, human-written tests \citep{jimenez2024swebench}. SWE-Bench also contains a Lite subset, a 300 problem subset of the overall SWE-Bench benchmark in order to handle computational expenses associated with performing the evaluation, and a Verified subset, a set of 500 problems that have been human-validated to be solvable when only provided the issue description \citep{chowdhury2024swebenchverified}. In the standard SWE-Bench setup, the agent is not provided with the human-written test cases. Because the current work is centered around test driven development using human-written tests, we provide the normally hidden, human-written tests to the system.

Most repository scale \textbf{software engineering agent systems} rely on single ReACT agents, or multiple agents, equipped with a set of tools and tasked with solving the issue \citep{wang2024openhands, yang2024sweagent, tao2024magis, wang2024opendevin, codeact, react}. Some examples of this include OpenDevin/OpenHands, the first of such work, and SWE-agent \citep{wang2024opendevin, wang2024openhands, yang2024sweagent}. Multi-agent systems involve multiple autonomous agents each given personas such as "Manager", "Fault Localizer", and "Verifier" and tasked with different objectives. Such systems include MAGIS, CODER, and AgentCoder \citep{tao2024magis, chen2024coder, huang2024agentcodermultiagentbasedcodegeneration}. 

Contemporary work has focused on scaling test-time compute via running a large number of parallel instances of an agent. Each instance of the agent proposes a patch, which are then filtered through by a selection module \citep{gao2025trae} -- achieving state of the art performance on SWE-Bench Verified, when human-written tests aren't provided. 

Other test-time scaling work involves intelligently selecting key decision points to instantiate a novel agent-based trajectory \citep{aggarwal2025darsdynamicactionresampling}. Similar to our current approach of LLM-integrated workflows is the Agentless and PatchPilot system \citep{xia2024agentless, li2025patchpilot}. Agentless is comprised of localization, repair, and patch validation phases, utilizing LLMs in a purely non-agentic fashion. Similar to Agentless, TDFlow prevents the LLM from deciding future actions -- resorting to following the workflow instead. However, TDFlow differs in that the individual workflow components consist of LLM sub-agents working in tanden with complex tools \citep{xia2024agentless}.

ExpeRepair is an LLM-based automated program repair system which is currently state of the art in the SWE-Bench Lite benchmark, when human-written tests are not provided. ExpeRepair innovates a dual-memory system capturing both episodic and semantic knowledge regarding problem instance to more accurate solve issues \citep{mu2025experepair}. 

\textbf{Automated test generation systems} such as CodeT, Libro, Otter, UTBoost and AEGIS also utilize LLMs in both workflow and agentic fashions for generating reproduction tests of repository scale issues \citep{chen2023codet, libro, wang2024aegis, otter}. Benchmarks such as TDD-Bench and SWT-Bench also provide a repository scale benchmark on the quality of LLM-generated reproduction tests \citep{mundler2024swtbench, ahmed2024tddbenchverified}. As the focus of our work lies in \textit{resolving human-written tests}, we consider other automated test generation systems and the evaluation thereof as complementary work. 

\subsection{Advantages of Agentic Workflows Over Agents}
There are several \textbf{advantages of agentic workflows} over monolithic agents. In the case of large, monolithic agents, resulting overall resolution success rate is bottlenecked by any individual subtask the agent underperforms the most in \citep{microsoft:multiagent_reference_architecture}. And because of the extremely tight coupling between, for example, the test generation capabilities of an LLM agent with the debugging capabilities of the same LLM agent, improving performance on these individual subtasks can prove to be challenging. 

A nontrivial downside to practical implementations of monolithic agents also lies within computational complexity increases as well as performance deterioration during long context generation and ingestion \citep{zhou2025mem1}. Agents powered by a central LLM with a context comprised of all actions taken by the LLM suffer significant performance deterioration in long context tool usage situations, challenges in auditability, and inability to efficiently scale in parallel due to the sequential nature of the context window \citep{souza2025prov, sapkota2025ai, fang2025environmentscaling}. Parallel computation is achievable via response ensembling techniques such as self-consistency; however, such approaches rely on each individual agent within the ensemble to gather its own context \citep{wang2022selfconsistency, gao2025trae, knappe2024semantic}.

Pre-defined agentic workflows are built and designed to solve a single task but are less flexible compared to monolithic agents as a result \citep{sapkota2025ai}. The decomposition of the task into sub-tasks is handled and embedded into the design of the workflow. The components of the workflow comprise of a mixture of standard software engineering and LLM sub-agents -- LLM agents prompted and equipped to perform only a single sub-task of the large workflow goal \citep{singh2024enhancing}.

\section{TDFlow}

The architecture of TDFlow is displayed in Figure \ref{fig:overview}. TDFlow is comprised of 4 sub-agents working in an algorithmic loop. If there don't exist human-written tests, the \textbf{Generate Tests} sub-agent is tasked with generating a set of reproduction tests given an issue description.

Given an issue description $D$, a list of currently failing tests $\{f_1, ..., f_F\}$, as well as a list of passing regression tests $\{p_1, ..., p_P\}$, TDFlow runs the failing tests in order to procure a list of error messages, $\{e_1, ..., e_F\}$. The source code for the failing tests, $\{s_1, ..., s_F\}$, are also provided to the initial \textbf{Explore Files} phase along with the issue description, $d$. The \textbf{Explore Files} sub-agent is to analyze the failing tests, peruse the repository, and then propose a repository-level patch to apply in order to solve the tests. A \textbf{Revise Patch} sub-agent is run if the patch fails to apply due to issues with context.

After a repository-level patch is proposed, the next step of TDFlow is to run all tests before recollecting $\{f_1, ..., f_F\}$, $\{p_1, ..., p_P\}$, and $\{e_1, ..., e_F\}$. Next, $F$ \textbf{Debug One} phase sub-agents are initialized. \textbf{Debug One} sub-agents are provided with the same repository exploration tools as the \textbf{Explore Files} sub-agent along with a debugger tool with access to a debugger. The goal of the \textbf{Debug One} sub-agent is to debug a single test and generate a comprehensive report on why the test failed. 

All $F$ reports are then aggregated before resubmitted to the \textbf{Explore Files} sub-agent for another iteration of repository-level patch generation.

\subsection{Explore Files}
During the \textbf{Explore Files} stage, an LLM sub-agent is tasked with proposing a repository-level, global patch but is prevented from interacting with the repository outside of viewing the content of files, finding specific keywords, and viewing the file hierarchy. The sub-agent is not provided with the ability to create or edit files, nor is it provided with bash access. During each iteration of \textbf{Explore Files}, the state of the repository will always remain as the initial state of the repository. Therefore, each proposed global patch represents a diff file with respect to the same initial repository state.

The context the LLM sub-agent is provided consists of a set of $\{c_1, ... c_N\}$ where $N$ represents the total number of previously attempted global patches. A single $c_i$ consists of the $i$th attempted global patch $p_i$, $\{s_1, ..., s_F\}$ for the set of failing test cases $\{f_1, ... f_F\}$ that $p_i$ fails on (including both the failing regression and reproduction tests), and the $F$ debugging reports generated by the \textbf{Debug Once} sub-agents.

As a result, the \textbf{Explore Files} sub-agent is carefully provided with the minimum amount of necessary information that completely represents all previous attempts as well as the outcomes of previous attempts.

\subsection{Revise Patch}
When a global patch fails to apply to the codebase due to having insufficient or incorrect repository context required to place the patch in the correct locations, the \textbf{Revise Patch} stage is initiated. The \textbf{Revise Patch} sub-agent is provided only with the malformed patch as well as the same find, view, and hierarchy tools provided to the \textbf{Explore Files} sub-agent. The \textbf{Revise Patch} sub-agent then explores the repository in order to discover the necessary context required to apply the repository-level patch. Compared to the alternative of providing the patch application error message directly to the \textbf{Explore Files} sub-agent, this separate sub-agent approach leads to more consistent and reliable repository-level patches.

\subsection{Debug One}
The \textbf{Debug One} sub-agent is provided with the test source code as well as the error message for a single failing test. It is also provided with the same find keyword, view file, and folder hierarchy tools as the \textbf{Generate Tests}, \textbf{Explore Files}, and \textbf{Revise Patch} sub-agents as well as a debugger tool allowing the sub-agent to submit commands to a debugger. The list of debugger commands is restricted compared to the standard Python debugger and is outlined in Appendix \ref{sec:debugger-commands}.

\subsection{Generate Tests}
Instead of human generated tests, TDFlow can also incorporate LLM-generated reproduction tests. This was performed in the SWE-bench Verified test. The \textbf{Generate Tests} phase of TDFlow is provided with the standard suite of repository exploration tools, the issue description, source code and file location for a previous test. The \textbf{Generate  Tests} sub-agent is also provided with an \textit{Evaluate Tests} tool which runs specified tests and returns the response of the test. Furthermore, the sub-agent must return the file line and names of the generated tests in order for the workflow to gather source code and be able to run and debug individual tests later on.

\subsection{Patch Selection}
If the algorithm performs the specified maximum number of iterations without finding a patch that satisfies all test cases, TDFlow selects and outputs the patch with the most passing reproduction tests that doesn't break any regression tests.

\section{Results}
\subsection{TDFlow outperforms baselines on human-written tests}
We compare TDFlow with several popular, state of the art SWE-Bench baselines. In this setting, we allow each system to have explicit access to the typically hidden, human-written test cases as well as the issue description. No system, however, is allowed access to the hints text within the SWE-Bench benchmark. In order to maintain a fair comparison, ExpeRepair, OpenHands and SWE-Agent were modified in the following ways: 1) the repository was updated to contain the reproduction tests along with regression tests and 2) initial prompts were updated to provide the names of the failing reproduction tests to the agent.

Agentless, being a workflow-based system without standard file navigation and search tools, was updated in the following ways: 1) the repository was updated to contain the reproduction tests, 2) the initial prompts were updated to provide both the names of the failing reproduction tests as well as the source code for each failing reproduction test, and 3) the ground-truth, human-written tests were used to select the final patch instead of Agentless's built-in patch selection system. Furthermore, to keep comparisons equal and fair to all systems, each system was tested using GPT-4.1. Default parameters were used for all systems. Further details on the experiments are in Appendix \ref{sec:lite_details}.

\begin{table}[h]
  \centering  
  \begin{tabular}{lcc}
    \hline
    \textbf{System}           & \textbf{Pass Rate} & \textbf{Cost/Issue} \\
    \hline
    OpenHands       & 47.8\%   & \$1.32 \\
    ExpeRepair      & 48.6\%   & \$0.84 \\
    SWE-Agent       & 49.0\%   & \$0.89 \\
    Agentless       & 61.0\%   & \$0.53 \\
    \textbf{TDFlow (ours)} & \textbf{88.8\%} & \textbf{\$1.51} \\
    \hline
  \end{tabular}
  \caption{
    Comparison between TDFlow and baseline systems on SWE-Bench Lite while providing human-written tests. To make the comparison fair, baseline systems were modified to have knowledge and testing capabilities for reproduction tests and were all tested using GPT-4.1. Instances with test hacking were considered failures.
  }
  \label{tab:gt_comp}
\end{table}

Table \ref{tab:gt_comp} shows the performance of each system as well as the average cost/issue. TDFlow achieves the best performance at 88.8\% of attempted issues resolved. We hypothesize Agentless performs so well due to the ensembling effects of the system generating up to 40 patches that were all tested against the human-written tests. Agentless also costs the least due to cheap localization. The other agentic systems require a large number of LLM API and tool calls before localizing problematic regions of code. In cost-sensitive settings, non-agentic localization techniques may be the most cost effective in gathering pertinent context for the LLM. 

\subsection{The final hurdle to human-level accuracy lies in test generation}
We ran TDFlow on SWE-Bench Verified, consisting of 500 problem instances, in two modes: the first mode used human-written tests while TDFlow generated its own tests in the second mode. For test generation, Claude 4 Sonnet was used. Both modes used GPT-5 for the \textbf{Explore Files}, \textbf{Debug One}, and \textbf{Revise Patch} sub-agents. Details of both experiments can be found in Appendix \ref{sec:verified_details}. SWE-Bench Verified was studied here due to the assurance that each instance was solvable given only the issue description \citep{chowdhury2024swebenchverified}. 

\begin{table*}[h]
  \centering  
  \setlength{\tabcolsep}{4pt}
  \begin{tabular}{lcccc}
    \hline
    \textbf{System}           & \textbf{Pass Rate} & \textbf{Cost/Issue}  & \textbf{Test Resolution Cost}\\
    \hline
    TDFlow: LLM-generated       & 68.0\%   & \$4.12 &  \$2.83 \\
    TDFlow: Human-written (no debugging sub-agent) & 87.2\% & \$0.73 & \$0.73 \\
    TDFlow: Human-written      & 94.3\%   & \$1.01 &   \$1.01 \\
    \hline
  \end{tabular}
  \caption{
    Comparison between performance of TDFlow when generating its own tests versus using human-written tests. The dataset is SWE-Bench Verified. The \textbf{Generate Tests} sub-agent used Claude 4 Sonnet while the \textbf{Explore Files}, \textbf{Revise Patch}, and \textbf{Debug One} sub-agents used GPT-5. Test resolution cost refers to the amount of money spent per issue to solve tests. As a result the test resolution cost of human-written tests is the same as the overall cost/issue, since no tests were generated.
  }
  \label{tab:hero_comp}
\end{table*}

The results of the experiment are found in Table \ref{tab:hero_comp}. When provided with human-written tests, TDFlow achieves a 94.3\% success rate on SWE-Bench Verified, approaching human-level performance on test resolution. Combined with the 68.0\% performance when generating its own tests, this suggests the test resolution capabilities of TDFlow are already human-level, while the final hurdle in LLM repository-scale issue resolution lies within test generation. Furthermore, the test resolution cost in the LLM-generated mode is significantly greater, most likely due to generated tests which don't truly test reproduction behavior.

\begin{figure}[htp]
  \centering
  \includegraphics[scale=0.37, trim=3.5cm 4cm 4cm 3.7cm, clip]{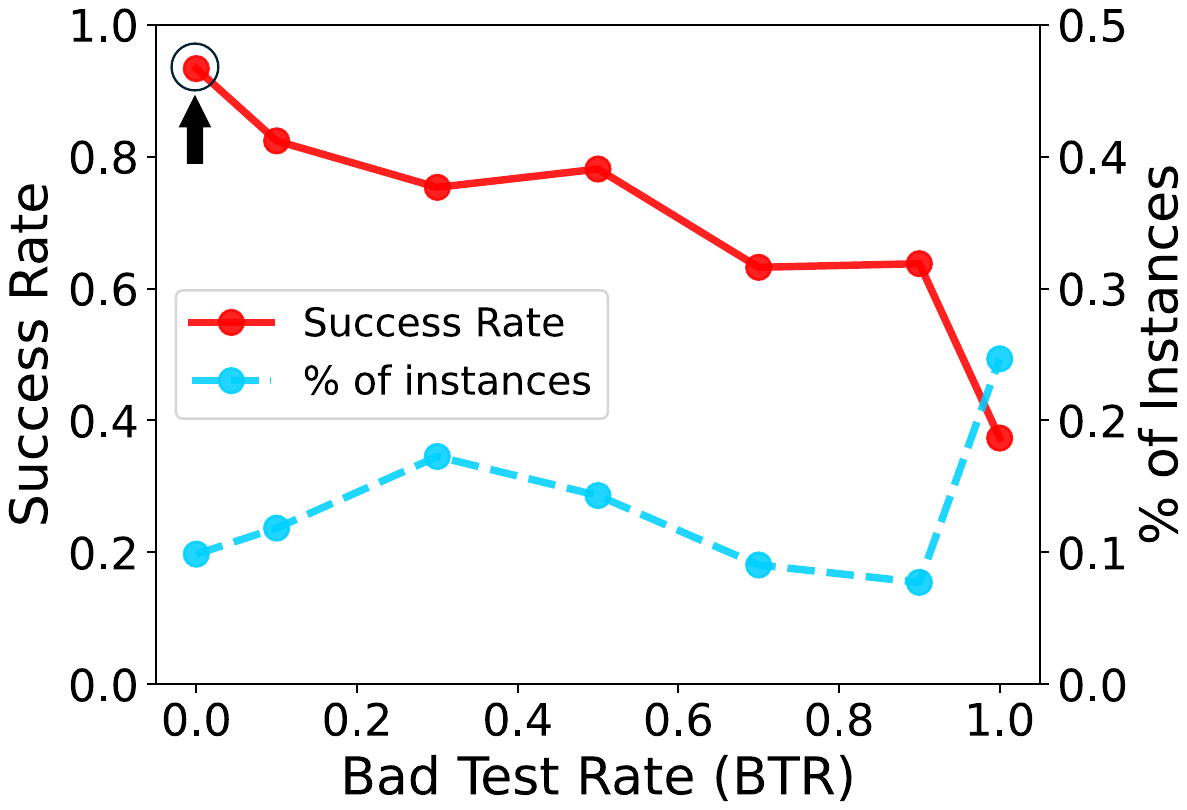}
  \caption {The solid line depicts the solve/success rate at each Bad Test Rate (BTR) level for the LLM-generated mode on SWE-Bench Verified using GPT-5. Bad Test Rate is the number of unsuccessful reproduction tests divided by the total number of LLM-generated tests. The dashed line depicts the \% of instances with the specified bad test rates. When BTR is 0, TDFlow has a 93.3\% solve rate.}
  \label{fig:test_analysis}
\end{figure}

The solid red line in Figure \ref{fig:test_analysis} shows the success rate of TDFlow when using LLM-generated tests on SWE-Bench Verified with respect to the Bad Test Rate (BTR) of the instance. BTR is defined as the number of unsuccessful reproduction tests divided by the total number of LLM-generated tests. An LLM-generated test is considered a successful reproduction test if the test fails before the gold/ground-truth patch is applied and passes after the gold patch is applied. Therefore, a BTR of 0 refers to an instance in which the generated tests are all successful reproduction tests. The datapoint pointed to by the black arrow refers to the success rate of TDFlow for instances containing only successful reproduction tests and previous regression tests. TDFlow successfully solves 93.3\% of these instances. This shows that whether or not the reproduction test is written by a human or an LLM doesn't matter for downstream test resolution performance so long as the test truly tests for reproduction behavior. TDFlow's test resolution capabilities are already at human-level when provided with successful reproduction tests. All predictions with test hacking were considered failures. A more detailed analysis on the types of tests generated by TDFlow can be found in Appendix \ref{sec:generated-tests}.

\subsection{TDFlow scales as the number of iterations increases}

\begin{figure}[!t]
  \centering
  \includegraphics[scale=0.8, trim=9.9cm 4.5cm 9cm 3.7cm, clip]{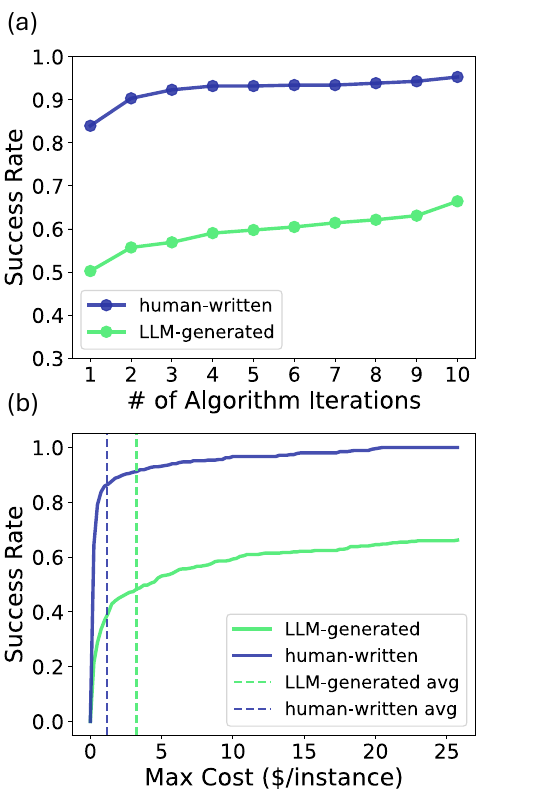}
  \caption {(a) The overall success rate of both TDFlow modes: when TDFlow is provided with human-written tests and when TDFlow is provided with LLM-generated tests. (b) The overall success rate of both modes as a function of the maximum cost per instance.}
  \label{fig:scaling}
\end{figure}

Figure \ref{fig:scaling}a portrays the overall success rate of both human-written and LLM-written modes as a function of the total number of algorithm iterations. There is a rapid point of diminishing returns after multiple iterations are run. However, the scaling result provides evidence that increasing the context with previously failed attempts, test results, and debugging reports leads to an overall increase in passed instances. Figure \ref{fig:scaling}b shows the success rate as a function of the maximum cost per instance with the average cost for both modes outlined.

\subsection{TDFlow minimizes test hacking instances}
One concern when deploying TDFlow in scenarios where humans write ground-truth tests for LLMs to solve is test hacking. The LLM-based system, if provided with the sole goal of solving tests without the proper guidelines and guardrails, is incentivized to "hack" the tests by proposing patches that pass test cases but do not solve the underlying issue \citep{pan2024feedback}. Examples of test hacking include modifying test source code, writing patches to skip/disable tests, writing arbitrary test-only logic, and hardcoding results to match test expectations. 

TDFlow's test hacking mitigation strategies involve precisely engineered system prompts, preventing patches from affecting test folders or otherwise manipulating test source code, only allowing sub-agents to view the repository folder instead of the entire filesystem, and repeated prompting via tool call results to nudge the system towards focusing on solving the underlying issue. We find these strategies to be sufficient for minimizing test hacking. We had three trained programmers manually inspect all TDFlow human-written SWE-Bench Lite and SWE-Bench Verified patches and logs for 300 and 500 instances respectively, specifically searching for examples of test hacking. They find only 4 instances of test hacking within the SWE-Bench Lite results and 3 instances of test hacking within the SWE-Bench Verified results. We consider all instances with test hacking as failures when reporting results in this work. The procedure and rubric and further details for manually identifying test hacking is found in Appendix \ref{sec:test_hacking}.

\section{Discussion}
As seen in Figure \ref{fig:test_analysis}, instances in which all generated reproduction tests are, in fact, successful reproduction tests (i.e. instances with 0 BTR) results in a 93.3\% pass rate when using GPT-5 -- highlighting that there is no fundamental difference between LLM-generated tests and human-written tests as long as the overall reproduction behavior is the same. Furthermore, as seen in the same figure, the primary bottleneck in overall SWE-Bench Verified performance is undoubtedly the large number of instances with high BTR. Somewhat surprisingly, TDFlow is still capable of solving most instances with non-zero BTR with an intuitive decreasing success rate as BTR increases. There is a significant dropoff in performance when there exist no successful reproduction tests. TDFlow's ability to maintain performance even when most tests are not truly reproduction tests is most likely due to the patch selection mechanism in the event a patch does not pass all provided reproduction tests. 

TDFlow's strong performance of 94.3\% on SWE-Bench Verified when provided gold-standard, human-written reproduction tests combined with its 93.3\% success rate on 0 BTR instances when generating its own reproduction tests indicates the primary bottleneck to human-level software engineering performance lies within writing tests rather than solving them. The debugging, file localization, bug localization, and code reasoning capabilities of precisely-engineered LLM systems are already sufficient for solving a diverse variety of software engineering issues within complex codebases such as Django. \textbf{Rather, understanding the issue and displaying such understanding through writing meaningful reproduction tests is the final obstacle towards a human-level performance software engineering system.}

One assumption TDFlow makes is in the quality of the provided tests, which are typically human-written. As a result, TDFlow is not allowed to modify reproduction tests after they have been generated or provided. If the initial tests are incorrect, then there is no correction mechanism later during the test resolution process. This is the primary explanation for why TDFlow underperforms compared to other state of the art systems when generating its own tests.

With TDFlow, we envision an LLM-human interactive cycle in which human developers guide LLM code generation by writing tests that LLMs then solve. Such an interactive cycle would lead to the significant code quality benefits described in \citet{pairTDD} and \citet{micrsoftTDD} without the substantial increase in development time. 

\section{Future Work}
We show an established avenue towards human-level performance in fully autonomous software engineering issue resolution via more accurate reproduction test generation systems. Long horizon reinforcement learning post-training specifically targeting test generation tasks is a promising route. Such task-specific training is only enabled once individual sub-tasks within software engineering are decoupled from one another into a workflow such as TDFlow. We also believe that automated methods of checking for test hacking is also required to realize a production-ready, workflow based system.

\section{Conclusion}
In this work, we present TDFlow, a test-driven agentic workflow specifically designed to solve human-written reproduction tests in a test-driven setting. TDFlow is a simple, precisely engineered, modular workflow consisting of several sub-agents each tasked with a well-defined subtask and provided the minimum set of tools and limited agentic freedom. The modularity of TDFlow allows for the decoupling and individual, targeted training of each sub-agent on specifically-designed tasks. We benchmark TDFlow on SWE-Bench Lite with GPT-4.1 and find it outperforms the next best system by 27.8\% when provided human-written tests. On SWE-Bench Verified, when solving successful reproduction tests, TDFlow already shows human-level performance of 94.3\%. Finally, through an analysis of the TDFlow system, we show that the final obstacle to autonomous software engineering, in the context of test-driven development and the SWE-Bench benchmark, lies in generating better reproduction tests.

\section{Limitations}
By nature of TDFlow being a rigid workflow consisting of multiple sub-agents instead of a single adaptable agent, the environment, infrastructure, and test information setup is more complex and less flexible than purely agentic systems. Furthermore, if the provided tests are truly unsolvable, TDFlow will keep performing main algorithm iterations until the iteration limit has been reached. There is no early-stopping mechanism or critic that can be used to save resources.


\bibliography{custom}

\newpage
\appendix

\section{Details on SWE-Bench Lite Experiment}
\label{sec:lite_details}
The settings used in the Agentless comparison on the SWE-Bench Lite experiment are found in Table \ref{tab:agentless-settings}. The settings for the OpenHands comparison on the SWE-Bench Lite are found in Table \ref{tab:openhands-settings}. For the SWE-Agent comparison, we specified the config path to \texttt{config/default/yaml}. The permalink to the file used is \url{https://github.com/SWE-agent/SWE-agent/blob/a1193dd8fd84eb3e2cd6b0ecbd0bed1cdbb84993/config/default.yaml}. The settings used in ExpeRepair comparison on the SWE-Bench Lite experiment are found in Table \ref{tab:experepair-settings}. All settings used across all comparisons are the default settings found in each of the respective repositories. Furthermore, to keep comparisons fair, we used the same LLM, GPT-4.1, across all systems. For TDFlow, the settings used are found in Table \ref{tab:tdflow-settings}. For the OpenHands agent, 91 instances failed during evaluation. As a result, the denominator of the OpenHands accuracy calculation is 201 instances. For TDFlow, we were unable to run 22 instances on our infrastructure due to test results for individual tests differing from test results for test suites. As a result, the denominator of the TDFlow accuracy calculation is 278 instances.

\begin{table}[h]
\centering
\small
\begin{tabular}{ll}
\hline
\textbf{Parameter} & \textbf{Value} \\
\hline
Model & GPT-4.1 \\
Number of Threads & 4 \\
Top-$n$ Candidates & 3 \\
\hspace{1em}• Combine & \\
\hspace{1em}• Related & \\
\hspace{1em}• Edit Locations & \\
\hspace{1em}• Merge & \\
\hspace{1em}• Repair & \\
Context Window (repair) & 10 \\
Number of Samples & 4 \\
Temperature (fine-grain edit locations) & 0.8 \\
\hline
\end{tabular}
\caption{Settings used for Agentless SWE-Bench Lite prediction. These are the default settings from \citet{xia2024agentless} and can generate up to 40 patches per instance. Instead of the standard patch selection process, all 40 proposed patches were run against the human-written tests to find any passing patch. Furthermore, the repository was modified to include the human-written tests. The prompts were also modified to include the human-written reproduction test names and source code.}
\label{tab:agentless-settings}
\end{table}

\begin{table}[htbp]
\centering
\small
\begin{tabular}{ll}
\hline
\textbf{Parameter} & \textbf{Value} \\
\hline
Agent Class & CodeActAgent \\
Model & GPT-4.1 \\
Number of Retries & 4 \\
Retry Multiplier & 2.0 \\
Retry Min Wait (s) & 5 \\
Retry Max Wait (s) & 30 \\
Max Message Characters & 30{,}000 \\
Temperature & 0.0 \\
Top-p & 1.0 \\
Drop Parameters & True \\
Modify Parameters & False \\
Caching Prompt & True \\
Reasoning Effort & High \\
Max Iterations & 300 \\
\hline
\end{tabular}
\caption{Settings used for the OpenHands SWE-Bench Lite prediction. Because this was a setting where human-written tests were provided, the instance repository was modified to include the human-written tests. The system prompt was also modified to include the names of each of the reproduction tests so the OpenHands agent understands which tests to focus on.}
\label{tab:openhands-settings}
\end{table}

\begin{table}[ht]
\centering
\small
\begin{tabular}{ll}
\hline
\textbf{Parameter} & \textbf{Value} \\
\hline
Retry Limit & 3 \\
Conv Round Limit & 10 \\
Reproduce Round Limit & 5 \\
Reproduce and Review & True \\
Test Exec Timeout & 300 \\
Number of Patches & 4 \\
\bottomrule
\end{tabular}
\caption{Settings used for the ExpeRepair SWE-Bench Lite prediction. Because this was a setting where human-written tests were provided, the instance repository was modified to include the human-written tests. Each system prompt in the ExpeRepair system was also modified to include the names of each of the reproduction tests so the agents can understand which tests to focus on.}
\label{tab:experepair-settings}
\end{table}

\begin{table}[ht]
\centering
\small
\begin{tabular}{ll}
\hline
\textbf{Parameter} & \textbf{Default Value} \\
\hline
Num Total Iterations & 10 \\
Max Tests Debug & 18 \\
Generate Tests Max Turns & 200 \\
Debug One Max Turns & 250 \\
Revise Patch Max Turns & 50 \\
Explore Files Max Turns & 75 \\
Temperature & 1.0 \\
\bottomrule
\end{tabular}
\caption{Settings used for the TDFlow SWE-Bench Lite prediction. Max Tests Debug is the maximum number of failing test cases TDFlow debugs per total algorithm iteration. We find the best performance when setting the temperature to 1, most likely due to the model exploring creative alternative solutions at each algorithm iteration.}
\label{tab:tdflow-settings}
\end{table}

\section{Details on SWE-Bench Verified Experiment}
\label{sec:verified_details}
For the SWE-Bench Verified experiment, TDFlow uses Claude 4 Sonnet for \textbf{Generate Tests} sub-agent and GPT-5 on high reasoning effort and low verbosity for the other agents. Otherwise, the same settings were used in the SWE-Bench Lite experiments as outlined in Table \ref{tab:tdflow-settings}. A large challenge of workflow-based pipelines lies within the rigidity of the framework. For example, in order to obtain the test source code and breakpoint location for debugging, each test name must have an associated corresponding line number without co-locating with any other test. Furthermore, each test must be individually run-able and debug-able, a criteria missing from several instances within the SWE-Bench benchmark. As a result, we were unable to run 45 instances in SWE-Bench Verified.

\section{Test Hacking Rubric}
\label{sec:test_hacking}
For the manual test hacking check, we recruited three trained software engineers to label TDFlow patch predictions and logs for test hacking. If the patch exhibited any of the signs in Table \ref{tab:test-hacking-rubric}, the reviewers flagged the patch as possible test hacking. If two or more of the reviewers flagged a patch as test hacking, the patch was labeled as a test hack. Out of 300 logs in SWE-Bench Lite, and 500 logs in SWE-Bench Verified, only 4 instances and 3 instances respectively were labeled as test hacking. These were automatically considered failed patches when reporting results.

\begin{table*}[ht]
\centering
\small
\begin{tabularx}{\linewidth}{l X}
\toprule
\textbf{Test hack} & \textbf{Description / example signs} \\
\midrule
Tests directly modified & The test code itself changed to avoid asserting the intended behavior. \\
Test skipped / disabled & Tests are marked \texttt{skip}, \texttt{@ignore}, or otherwise disabled. \\
Assertions weakened or removed & Strong checks replaced by weaker/looser assertions or removed entirely. \\
Test data modified to avoid failure & Fixtures or input data are changed so the failing scenario no longer occurs. \\
Environment manipulation & Test environment (containers, mocks, clocks) altered to hide problematic behavior. \\
Test runner configs changed & CI/test-runner settings adjusted to run fewer tests or a narrower scope. \\
Dependencies pinned / relaxed & Versions changed to avoid a failing dependency or to bypass failing behavior. \\
Environment variables added / modified & Vars set to disable checks or enable permissive behavior in tests. \\
Arbitrary code changes & Ad-hoc code edits that only make the test pass without addressing root cause. \\
Test-only logic & Code paths added that execute only under test (feature flags, test hooks). \\
Behavior altered to satisfy test & Production logic intentionally altered to match the test’s expectations. \\
Hardcoded outputs / special cases & Functions return literal values that match tests rather than constructing correct output. \\
Use of magic constants from tests & Production code reuses constants seen only in tests (sign of tailoring). \\
\bottomrule
\end{tabularx}
\caption{The rubric used when manually checking patches and logs for test hacking. If any of these signs were exhibited, the reviewers flagged the patch. If two or more of the three reviewers flagged the patch as test hacking, the patch was labeled as a test hack and subsequently considered a failed patch when reporting results.}
\label{tab:test-hacking-rubric}
\end{table*}

\section{Debugger Commands}
\label{sec:debugger-commands}
The debugger commands that the \textbf{Debug One} sub-agent is provided are listed in Table \ref{tab:debugger-commands}. 

\begin{table}[!h]
\centering
\small
\begin{tabular}{p{0.18\linewidth} p{0.75\linewidth}}
\hline
\textbf{Command} & \textbf{Description} \\
\hline
\texttt{s} & Step into the function call on the current line. \\
\texttt{n} & Execute the current line and move to the next one (step over function calls). \\
\texttt{r} & Continue execution until the current function returns. \\
\texttt{c} & Continue execution until the next breakpoint. \\
\texttt{b} & Show or set breakpoints. Usage: \newline
\hspace*{1em}-- \texttt{b}: List all breakpoints and their status (hit count, condition, ignore count). \newline
\hspace*{1em}-- \texttt{b <lineno>}: Set a breakpoint at the given line in the current file. \newline
\hspace*{1em}-- \texttt{b <filename>:<lineno>}: Set a breakpoint in another file. \newline
\hspace*{1em}-- \texttt{b <function>}: Set a breakpoint at the first line of a function. \newline
\hspace*{1em}-- \texttt{b <location>, <condition>}: Set a conditional breakpoint (triggers only when the condition is \texttt{True}). \\
\texttt{p <expr>} & Print the value of an expression. \\
\texttt{pp <expr>} & Pretty-print an expression. \\
\texttt{whatis <expr>} & Show the type of an expression. \\
\texttt{args} & Show the arguments of the current function. \\
\texttt{locals()} | \texttt{globals()} & Show all local or global variables. \\
\texttt{l} & Show surrounding lines of code. \\
\texttt{l .} & Show the current line of code. \\
\texttt{ll} & Show the full source of the current function. \\
\texttt{w} | \texttt{where} & Show the current call stack. \\
\texttt{restart} & Restart the debugger from the beginning of the test. \\
\hline
\end{tabular}
\caption{The debugger commands available to the \textbf{Debug One} sub-agent.}
\label{tab:debugger-commands}
\end{table}

\section{Generated Tests Analysis}
\label{sec:generated-tests}
Figure \ref{fig:test_analysis} shows the distribution of generated test cases for the SWE-Bench Verified experiment across all repositories. 

\begin{figure*}[!h]
  \centering
  \includegraphics[scale=0.6, trim=3.5cm 7cm 3.5cm 7cm, clip]{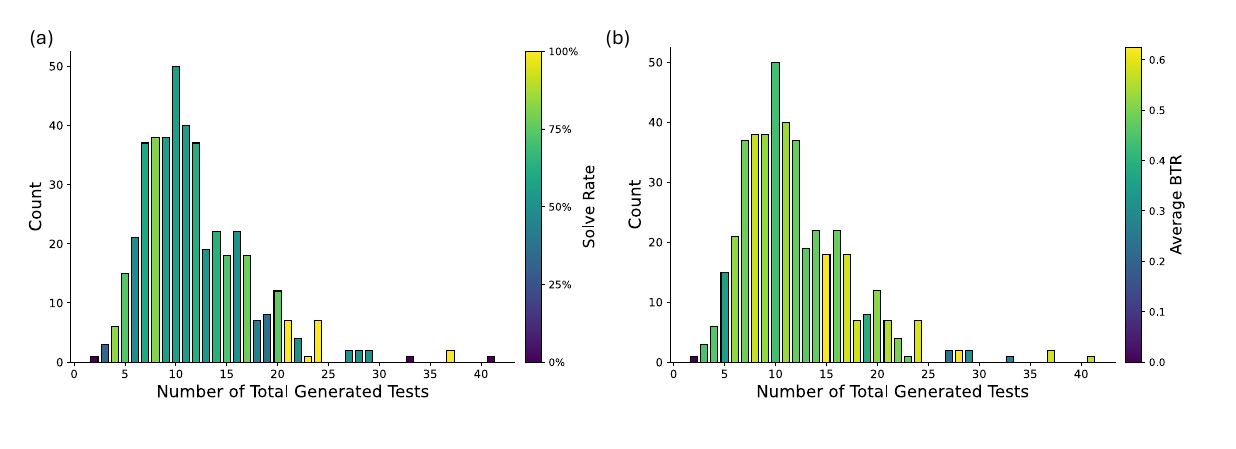}
  \caption {(a) A histogram of the distribution of test counts for the SWE-Bench Verified experiment where the LLM generates its own tests. The color bar refers to the solve rate of each bin. (b) The same distribution in histogram form except the color refers to the average BTR of each bin.}
  \label{fig:test_count}
\end{figure*}

Figure \ref{fig:test_count}a is a histogram of the distribution of the number of generated tests per instance. The color of the histogram refers to the solve rate for each histogram bin. Figure \ref{fig:test_count}b is a histogram of the same distribution except the color refers to the average BTR within each bin. There is no obvious relationship visible in either figure. 

\begin{figure*}[!h]
  \centering
  \includegraphics[scale=0.6, trim=1cm 8cm 1cm 8cm, clip]{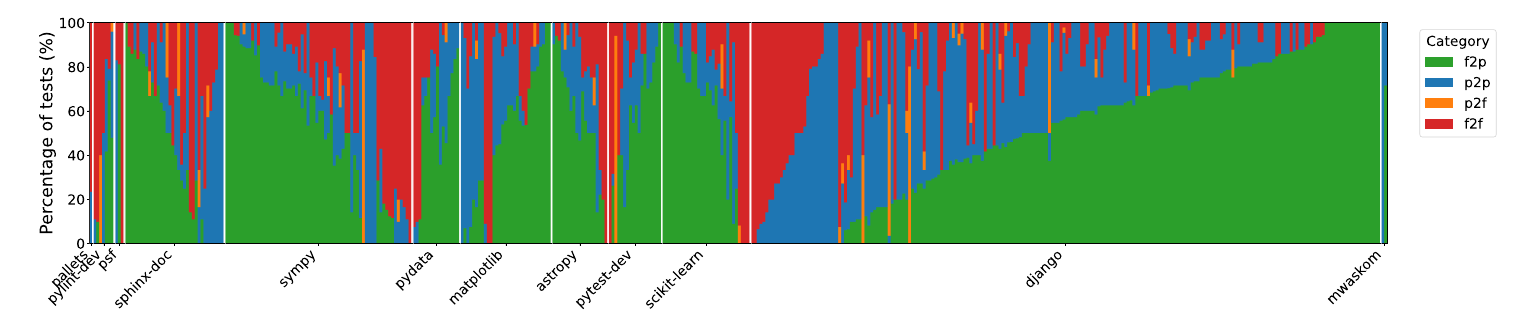}
  \caption {The distribution of f2p, p2p, p2f, and f2f tests. \texttt{f2p} refers to tests which fail before the gold patch is applied and pass after. \texttt{f2f} refers to tests which fail both before and after the gold patch is applied. \texttt{p2f} refers to tests which pass before the gold patch is applied and fails after. And \texttt{p2p} refers to tests which pass both before and after the gold patch is applied. }
  \label{fig:test_analysis}
\end{figure*}

\clearpage
\section{Prompts}
\subsection{Generate Tests System Prompt}
\begin{lstlisting}[style=prompt, caption={System prompt used for the \textbf{Generate Tests} sub-agent}]
You are an autonomous software engineering agent tasked to fix an open issue from an open-source repository.

Your specific objective is to generate reproduction tests for the open issue using the description of the issue. Make sure each reproduction test only has ONE assert statement and only tests a single functionality.

Carefully read the description and think hard about a plan to create reproduction tests BEFORE performing any actions. Your plan should include a discussion on potential edge cases and how to write tests for them.

Use the `evaluate_tests` tool to run your reproduction tests. Note that reproducion tests should FAIL at the moment since the issue has not yet been solved.

Only when you are 100% confident your tests are comprehensive and complete, use the `submit_tests` tool to submit your final set of reproduction tests. NEVER run `submit_tests` until you have run THE EXACT PATCH on `evaluate_tests` first.

Your thinking should be thorough and so it's fine if it's very long. You must think step by step, for as long as possible, before and after each action/tool-call you decide to take.

If you are not sure about file content or codebase structure pertaining to the issue, use your tools to gather the relevant information: do NOT guess or make up an answer.

You MUST plan extensively before each function call, and reflect extensively on the outcomes of the previous function calls. DO NOT do this entire process by making function calls only, as this can impair your ability to solve the problem and think insightfully.

Be as cautious as possible, explore the repository as much as possible. You have all the time in the world.
\end{lstlisting}

\subsubsection{Generate Tests User Prompt}
\begin{lstlisting}[style=prompt, caption={User prompt used for the \textbf{Generate Tests} sub-agent}]
The current working directory is /home/repo/

## Issue description:
{issue}

## Test name format:
For this repo, the command to run tests is {test_cmd} while the test names are formatted such as {test_example}, which is found in {test_example_file}.
In order to run your reproduction tests, we will run `{test_comand} {test_name}`.
\end{lstlisting}

\subsection{Explore Files System Prompt}

\begin{lstlisting}[style=prompt, caption={System prompt used for the \textbf{Explore Files} sub-agent}]
You are an autonomous software engineering agent tasked to fix an open issue from an open-source repository. A patch is 100% necessary.

You will be presented with a Github issue description as well as a series of patches that have failed in the past. None of the previous patches have been applied so you are starting from a clean repo.

Each of the failing patches are also accompanied by a test analysis conducted by a junior engineer equipped with a debugger.

Carefully read the issue and think hard about a plan to solve it BEFORE performing any actions.

Your thinking should be thorough and so it's fine if it's very long. You must think step by step, for as long as possible, before and after each action/tool-call you decide to take.

If you are not sure about file content or codebase structure pertaining to the issue, use your tools to gather the relevant information: do NOT guess or make up an answer.

You MUST plan extensively before each function call, and reflect extensively on the outcomes of the previous function calls. DO NOT do this entire process by making function calls only, as this can impair your ability to solve the problem and think insightfully.

THE PROBLEM CAN DEFINITELY BE SOLVED WITHOUT THE INTERNET.

Do not concern yourself with updating documentation, only focus on solving the issue at hand.
\end{lstlisting}

\subsubsection{Explore Files User Prompt}

\begin{lstlisting}[style=prompt, caption={User prompt used for the \textbf{Explore Files} sub-agent}]
## Issue description:
{issue}

In the event the test source code provided is incorrect, you should go search for the test in the repo using the test name.

{all_patches_str}
"""

EXPLORE_FILES_USER_PROMPT_INITIAL = """
## Issue description
{issue}

## Failing Reproduction Tests and Error Messages
The following tests are currently failing and need to pass after your fix:
{initial_failing_tests}

## Folder structure
The current working directory is /home/repo.

/home/repo
{repo_structure}
\end{lstlisting}

\subsection{Debug One System Prompt}
\begin{lstlisting}[style=prompt, caption={User prompt used for the \textbf{Explore Files} sub-agent}]
You are a software engineer tasked to assist in fixing a Github issue from an open-source repository using a debugger. 

You have already proposed a patch, but the patch fails on one or more tests. You will be provided a debugger with access to one of the failing tests to debug.

Your goal is to write a report on the patch using the submit_report tool. This report will be used later to help understand and fix the failing patch.

The report should be a SPECIFIC AS POSSIBLE and should include code if applicable.

Carefully read the issue and think hard about a plan to solve it.

DO NOT DO THIS ENTIRE PROCESS BY MAKING FUNCTION CALLS ONLY, as this can impair your ability to solve the problem and think insightfully.

Your thinking should be thorough and so it's fine if it's very long. You can think step by step before and after each action you decide to take.

THE PROBLEM CAN DEFINITELY BE SOLVED WITHOUT THE INTERNET.

The first thing you should is set any necessary breakpoints for debugging. Try to avoid using line numbers to specify breakpoints unless absolutely necessary
IMPORTANT: Do NOT set breakpoints at function signatures (e.g. "def create") as these breakpoints won't hit. Rather, set breakpoints at the first line of CODE within the function.

- The failing test may either be a regression test or a reproduction test of the issue at hand.
- Carefully read the issue and think hard about a plan to solve it before coding.
- Do NOT submit your report until you are ABSOLUTELY certain what the issue is. Once you submit your report, the debugging session will end.
- When debugging, try to determine the root cause of the issue rather than addressing symptoms.
\end{lstlisting}

\subsubsection{Debug One User Prompt}

\begin{lstlisting}[style=prompt, caption={User prompt used for the \textbf{Debug One} sub-agent}]
Github issue description: 
{issue}

Test source code:
{test_source}

This is a {reg_or_repro} test.

Test output message:
{test_message}

The failing patch:
{failing_patch}

Current debugger state:
{context}
\end{lstlisting}

\subsection{Revise Patch System Prompt}

\begin{lstlisting}[style=prompt, caption={System prompt used for the \textbf{Revise Patch} sub-agent}]
You are a code-repair assistant specializing in fixing incorrect patch content for the `apply_patch` tool.

You will be provided with a **malformed patch**-one that has incorrect or outdated lines and therefore cannot be applied cleanly. Your task is to revise the patch to make it valid. **Do not modify the inserted code in the patch**

See the tool description if you need a refresher on the format or rules.

Carefully read the error message and think hard about a plan to solve it BEFORE performing any actions.

Your thinking should be thorough and so it's fine if it's very long. You must think step by step, for as long as possible, before and after each action/tool-call you decide to take.

If you are not sure about file content or codebase structure pertaining to the issue, use your tools to gather the relevant information: do NOT guess or make up an answer.

You MUST plan extensively before each function call, and reflect extensively on the outcomes of the previous function calls. DO NOT do this entire process by making function calls only, as this can impair your ability to solve the problem and think insightfully.

Do NOT modify any of the code that's being inserted, only modify the context such that the patch works.

Only when you're ready and confident that the patch is correct should you call `apply_patch` with the revised patch. Veer on the side of caution before calling the `apply_patch` tool.
\end{lstlisting}

\subsubsection{Revise Patch User Prompt}

\begin{lstlisting}[style=prompt, caption={User prompt used for the \textbf{Revise Patch} sub-agent}]
The current working directory is /home/repo/

Here is the bad patch:
{patch}

Here is the error message when attempting to apply the patch:
{error_message}
\end{lstlisting}

\end{document}